%Paper: solv-int/9404002
%From: "R. A. Sharipov" <root@bgua.bashkiria.su>
%Date: Wed,  7 Apr 93 11:14:24 +0600
\input amstex
\input amsppt.sty
% Typeset by AMS-TeX version 2.1, IBM AT-386.
\topmatter
\title
Dynamical systems accepting the normal shift.
\endtitle
\author
Sharipov R.A.
\endauthor
\address
Department of Mathematics,
Bashkir State University, Frunze str. 32, 450074 Ufa,
Russia
\endaddress
\email
root\@bgua.bashkiria.su
\endemail
\dedicatory
Report on 16-th joint session of Seminar of I.G. Petrovsky\\
and Moscow Mathematical Society at Moscow State University.
\enddedicatory
\date
January 19, 1994.
\enddate
\endtopmatter
\document
     Newtonian dynamical system on a Riemannian manifold $M$ is a
system of ordinary differential equations of the form
$$
\dot x^i=v^i\hskip 10em\nabla_tv^i=F^i(\bold x,\bold v)
$$
describing the motion of a mass point with the unit mass in the
force field $\bold F$ on $M$. Let $S$ be the hypersurface in $M$
and let $\bold n(P)$ be the unit normal vector to $S$ at the
point $P$. Taking the initial velocity of unit mass points on
$S$ for $t=0$ as $\bold v=v(P)\bold n(P)$ we define the shift of
$S$ along the trajectories of the dynamical system: $f_t: S
\longrightarrow S_t$. The transformation $f_t$ is called the
normal shift if the trajectories of the dynamical system cross
the hypersurfaces $S_t$ along their normal vectors for any
value of $t$. \par
\definition{Definition} Newtonian dynamical system on $M$ is
called the system accepting the normal shift if for any
hypersurface $S$ one can find the function $v(P)$ on $S$ for the
modulus of initial velocity defining the normal shift of $S$.
That Newtonian system is called the strongly normal dynamical
system if the above function $v(P)$ can be normalized by the
condition $v(P_0)=v_0$ for any choice of $P_0\in S$ and for
any nonzero $v_0\in\Bbb R$.
\enddefinition
\proclaim{Theorem} The strong normality condition for the
Newtonian dynamical system is equivalent to the following system
of partial differential equations for its force field
$\bold F(\bold x,\bold v)$
$$
\left\{
\aligned
&(v^{-1}F_i+\tilde\nabla_i(F^kN_k))P^i_q=0\\
&(\nabla_i F_k+\nabla_k F_i-2v^{-2}F_i F_k)N^k P_q^i+
v^{-1}(\tilde \nabla_k F_i F^k-\tilde \nabla_k F^r N^k N_r F_i)
P_q^i=0\\
&P^k_iP^q_j\left(N^r\frac{\tilde\nabla_rF_k}{v}F_q
-\nabla_qF_k\right)=
 P^k_iP^q_j\left(N^r\frac{\tilde\nabla_rF_q}{v}F_k
-\nabla_kF_q\right)\\
&P^k_i\tilde\nabla_kF^qP^j_q=
\frac{P^k_r\tilde\nabla_kF^qP^r_q}{n-1}P^j_i
\endaligned
\right.
$$
where $N^i=|\bold v|^{-1}v^i$, $P^i_k=\delta^i_k-N_kN^i$ and the
covariant derivatives are defined as $\nabla_iF^k=\partial
F^k/\partial x^i+\Gamma^k_{ij} F^j-\partial F^k/\partial v^s
\Gamma^s_{ij} v^j$, $\tilde\nabla_iF^k=\partial F^k/
\partial v^i$.
\endproclaim
     The concept of dynamical systems accepting the normal shift
was introduced in \cite{1} and \cite{2} (see also \cite{3}).
Multidimensionlal dynamical systems accepting the normal shift
are considered in \cite{4} (see also \cite{5}). The above results
for the dynamical systems on Riemannian manifolds are published
in \cite{6} and \cite{7}. Their generalization for higher order
(non-Newtonian) dynamical systems is considered in \cite{8}.
\par
     Author thanks the International Scientific Fund of Soros
for financial support in 1993. \par

\enddocument